\documentclass[two column,superscriptaddress,aps,pra]{revtex4-1}

\usepackage{setspace}
\newcommand{\mrm}{\rm}

\newcommand{\udes}{ L_S }

\newcommand{\nb}{ {n^{}_B} }

\newcommand{\na}{{n^{}_A}}

\newcommand{\Is}{I_S}

\newcommand{\Ib}{I_B}

\newcommand{\btab}{\begin{tabular}}
\newcommand{\etab}{\end{tabular}}

\newcommand{\rank}{{\bf rank}}

\newcommand{\trace}{{\bf Tr}}

\newcommand{\un}{{\mathbf{1}}}

\newcommand{\real}{{\mrm Re}}
\newcommand{\imag}{{\mrm Im}}
\newcommand{\diag}{{\mrm diag}}

\newcommand{\gam}{\gamma}

\newcommand{\om}{\omega}

\newcommand{\del}{\delta}
\newcommand{\sig}{\sigma}

\newcommand{\Gam}{\Gamma}
\newcommand{\Om}{\Omega}

\newcommand{\norm}[1]{ \left\| #1 \right\| }

\newcommand{\rhob}{\bar{\rho}}

\newcommand{\Ccal}{{\mathcal C}}

\newcommand{\eg}{\emph{e.g.}}
\newcommand{\ie}{\emph{i.e.}}
\newcommand{\etc}{\emph{etc.}}

\newcommand{\bquem}{\begin{quote}\begin{em}}
\newcommand{\equem}{\end{em}\end{quote}}

\newcommand{\blist}{\begin{description}}
\newcommand{\elist}{\end{description}}

\newcommand{\bquote}{\begin{quote}}
\newcommand{\equote}{\end{quote}}

\newcommand{\ben}{\begin{enumerate}}
\newcommand{\een}{\end{enumerate}}

\newcommand{\bit}{\begin{itemize}}
\newcommand{\eit}{\end{itemize}}

\newcommand{\bea}{\begin{array}}
\newcommand{\eea}{\end{array}}

\newcommand{\bds}{\begin{displaystyle}}
\newcommand{\eds}{\end{displaystyle}}

\newcommand{\Rbf}{{\mathbf R}}
\newcommand{\Cbf}{{\mathbf C}}

\newcommand{\ds}{\displaystyle}

\newcommand{\refeq}[1]{(\ref{eq:#1})}

\newcommand{\refsec}[1]{\ref{sec:#1}}

\newcommand{\opt}{{\mrm opt}}

\makeatletter
\def\beq{\@ifnextchar 
[{\@tempswatrue\@beq}{\@tempswafalse\@beq[]}}
\def\@beq[#1]{\begin{equation}\edef\@tmparg{#1}\ifx\@tmparg\@e
mpty \else
	\label{#1}\fi}
\newcommand{\eeq}{\end{equation}}
\makeatother 

\newcommand{\beqaa}{\begin{eqnarray*}}
\newcommand{\eeqaa}{\end{eqnarray*}}

\newcommand{\beqa}{\begin{eqnarray}}
\newcommand{\eeqa}{\end{eqnarray}}

\newcommand{\bc}{\begin{center}}
\newcommand{\ec}{\end{center}}

\newcommand{\bfig}{\begin{figure}}
\newcommand{\efig}{\end{figure}}

\renewcommand{\Is}{I_n}

\renewcommand{\diag}{\mbox{\rm diag}}

\renewcommand{\na}{{n^{}_A}}

\renewcommand{\rank}{\mbox{\rm rank}}

\renewcommand{\udes}{U_S}

\newcommand{\phiopt}{\phi_{\rm opt}}
\newcommand{\Phiopt}{\Phi_{\rm opt}}
\newcommand{\Uobj}{U_{\rm obj}}
\newcommand{\U}{{\rm U}}
\newcommand{\SU}{{\rm SU}}

\renewcommand{\Ib}{I_{B}}
\renewcommand{\Is}{I_{S}}

\newcommand{\omb}{\bar{\om}}

\newcommand{\figref}[1]{Fig.~\ref{fig:#1}}

\renewcommand{\udes}{{W}}

\renewcommand{\na}{N_{A}}

\renewcommand{\nb}{{N_B}}

\newcommand{\nbsqr}{{N_B^2}}

\newcommand{\Kcal}{{\cal K}}

\newcommand{\Pcal}{{\cal P}}
\newcommand{\Vcala}{{\cal V}_{A}}
\newcommand{\Vcalb}{{\cal V}_{B}}

\newcommand{\Bcal}{{\cal B}}

\renewcommand{\rhob}{\rho_B}

\newcommand{\normfro}[1]{\norm{#1}_{\rm fro}}
\newcommand{\normnuc}[1]{\norm{#1}_{\rm nuc}}

\renewcommand{\Rbf}{\mathbb{R}}
\renewcommand{\Cbf}{\mathbb{C}}

\renewcommand{\trace}{\mbox{\bf Tr}}
\renewcommand{\rank}{\mbox{\bf rank }}

\newtheorem{thm}{Theorem}

\renewcommand{\un}{\mathbbm{1}}

\newcommand{\Hcal}{{\cal H}}

\newcommand{\Vcal}{{\cal V}}

\usepackage[colorlinks=true,
            urlcolor=blue,
            citecolor=red,
            pdfstartview=FitH,
            pdfpagemode=None]{hyperref}

\usepackage{times}
\usepackage{graphicx}
\usepackage{amssymb}
\usepackage{amsmath}
\usepackage{amsfonts}
\usepackage[dvips]{epsfig}

\usepackage{bbm}

\newcommand{\n}{{N}}
\renewcommand{\nb}{{N_B}}
\renewcommand{\na}{{N_A}}
\renewcommand{\nbsqr}{{N_B^2}}

\newcommand{\nasqr}{{N_A^2}}
\renewcommand{\U}{\text{U}}

\renewcommand{\Uobj}{{U_{\rm ext}}}

\begin{document}

\title{Quantum Control Landscape of Bipartite Systems}

\author{Robert L. Kosut}
\affiliation{
SC Solutions, Sunnyvale CA,
}
\author{Christian Arenz}
\author{Herschel Rabitz}
\affiliation{Princeton University, Princeton, NJ}


\begin{abstract}

The control landscape of a quantum system $A$ interacting with another
quantum system $B$ is studied. Only system $A$ is accessible through
time dependent controls, while system B is not accessible. The
objective is to find controls that implement a desired unitary
transformation on $A$, regardless of the evolution on $B$, at a
sufficiently large final time. The freedom in the evolution on $B$ is
used to define an \emph{extended control landscape} on which the
critical points are investigated in terms of kinematic and dynamic
gradients. A spectral decomposition of the corresponding extended
unitary system simplifies the landscape analysis which provides: (i) a
sufficient condition on the rank of the dynamic gradient of the
extended landscape that guarantees a trap free search for the final
time unitary matrix of system $A$, and (ii) a detailed decomposition
of the components of the overall dynamic gradient matrix.
Consequently, if the rank condition is satisfied, a gradient algorithm
will find the controls that implements the target unitary on system
$A$.  It is shown that even if the dynamic gradient with respect to
the controls alone is not full rank, the additional flexibility due to
the parameters that define the extended landscape still can allow for
the rank condition of the extended landscape to hold. Moreover,
satisfaction of the latter rank condition subsumes any assumptions
about controllability, reachability and control resources. Here
satisfaction of the rank condition is taken as an assumption. The
conditions which ensure that it holds remain an open research
question. We lend some numerical support with two common examples for
which the rank condition holds.

\end{abstract}

\maketitle

\section{Introduction}

Extensive theoretical and experimental evidence supports the relative
ease of searching for very good to near optimal controls for quantum
systems,
\cite{RabitzHR:04,RabitzHHKD:06,HsiehWRL:10,RussellRW:17}. The
``ease'' here refers to the search efficiency (\eg, number of
iterations), putting aside the overhead of
experimental/laboratory/field set up efforts. Similarly, in a
computational simulation context, numerical optimization algorithms
are typically very successful in finding classical control fields that
drive a quantum system towards a desired target. For instance,
maximizing the overlap $F(c(t))$ between a target state
(transformation) and the actual state (time evolution operator) of a
closed quantum system utilizing time dependent fields $c(t)$
\emph{almost always} succeeds using a gradient based search
\cite{RussellRW:17}. This finding suggests that the underlying
\emph{control landscape} \cite{RabitzHR:04} defined by the functional
$F(c(t))$ is ``simple'' in the sense that local optima (traps) rarely
occur. Thus, in the absence of traps, a gradient based search
typically converges to the global optimum, which, for instance,
corresponds to the preparation (implementation) of a target state
(gate) with fidelity $F=1$.  The scope of the principles underlying
this result seem to apply to more than quantum systems, with
applicability across a wide range of optimization in the sciences and
engineering in many highly complex systems \cite{RussellVR:18}.

However, any realistic quantum system interacts with its surrounding
environment causing the system dynamics to become perturbed in some
fashion. The search for optimal controls, and in particular the study
of the control landscape of such open quantum systems, is relatively
unexplored. A basic question is whether the interaction with the
environment, or with an auxiliary system in general, introduces traps
into the control landscape so that a gradient based search would not
succeed in finding the global optimum.

In this paper we investigate the control landscape of \emph{bipartite
  quantum systems}. That is, we study the optimization of a unitary
transformation objective for quantum system $A$ that is subject to
time dependent fields and coupled to another quantum system $B$, so
that the overall combined dynamics is unitary.  We derive a sufficient
criteria that allows for concluding when a gradient based algorithm
will be successful in finding the fields that implement a desired
unitary operation on system $A$.  In particular, using the fidelity
measure $F$ developed in \cite{GDKBH:10} for bipartite systems, we
find a sufficient condition for when $\nabla_{c} F\neq 0$ holds except
at saddles and the global maximum $F=1$ (and minimum), \ie, the
control landscape is \emph{trap-free}.  This result is achieved by
noting that, since we are only interested in implementing a unitary
operation on $A$, maximizing over the parameters describing the final
unitary operation on system $B$, collected in the vector $\phi$,
referred to as the extended landscape parameter, yields additional
freedom to aid in the optimization process.  The sufficient condition
is on the rank of the dynamic gradient of the spectral frequencies of
an associated extended unitary with respect to both $c$ and $\phi$.
Since $\phi$ becomes a global phase on the target unitary in the
absence of $B$, our results also hold for closed quantum systems.

In contrast to previous studies of the control landscape of quantum
systems, the trap-free condition presented here does not depend on the
assumption that the system is controllable, or on any assumption that
controls are available that allow for implementing the target unitary
transformation. As long as the developed sufficient condition holds, a
trap-free search is guaranteed. So as not to mislead the reader, it
remains an open challenge to find criteria for when the condition
holds. In this work satisfaction of the condition is stated as an
assumption, though its general validity is not known at this
time. Taking a pragmatic view, the rank condition can be used to test
available controls to lend support for their use.  In this regard we
provide numerical evidence that the condition is satisfied for many
simulations of two common bipartite systems; the results are fully
consistent with the main assumption and its consequences. The afore
stated sufficient condition follows directly by utilizing the spectral
decomposition of the extended landscape unitary matrix, which as a
by-product, also yields new detailed expressions for the relevant
gradients. These may provide the basis for establishing conditions
which underly the presented sufficient condition for a trap-free
search.

The work is organized as follows. We begin by introducing the
bipartite control system and the fidelity measure, followed by
defining the optimization problem. Using the spectral decomposition of
unitary operations we first show that the landscape can be expressed
as a function of what we refer to as the unitary ``frequencies''
$\omega(c,\phi)$ that depend on the controls $c$ and $\phi$.  An
immediate consequence is that the landscape as a function of $\omega$,
i.e, the so called kinematic landscape, only posses a finite number of
saddle points. By defining an \emph{extended control landscape} we
will see that if the rank of the gradient of $\omega$ with respect to
$c$ and $\phi$, \ie, the so called dynamic landscape, is full, then
the landscape of $A$ that arises from disregarding the evolution on
$B$ is trap-free. We numerically investigate two common models
describing a bipartite system and show that the rank condition holds
for the considered systems.

\section{Bipartite Quantum Control System}
We consider a closed \emph{bipartite} quantum system with parts $A$
and $B$ of dimension $N_{A}$ and $N_{B}$, respectively. The total
system of dimension $N=N_{A}N_{B}$ is described by a time independent
(drift) Hamiltonian $H_{0}$, which includes interactions between $A$
and $B$. We further assume that system $A$ can be influenced by $M$
time varying control fields $c_{m}(t),~m=1,\cdots M$ that are coupled
through $M$ (control) Hamiltonians $H_{m}$ in a bilinear way to system
$A$ \cite{elliott2009bilinear}. The time dependent Hamiltonian
describing the total system is then given by,
\begin{align}
\label{eq:totalH}
H(t)=H_{0}+\sum_{m=1}^{M}c_{m}(t)(H_{m}\otimes I_{N_{B}}),
\end{align} 
where $I_{N_{B}}$ denotes the identity matrix of dimension
$N_{B}\times N_{B}$.  We additionally assume that the controls are
piecewise constant over $L$ uniform intervals $\delta=T/L$. The total
unitary evolution denoted by $U(c)\in\U(\n)$ at time $t=T$ is then
given by a product of $L$ unitary operations. Specifically,
\beq[eq:uhc]
\bea{l}
U(c) = U_1\cdots U_L
\\
U_\ell = e^{-i\del H_\ell},\ \ell=1,\cdots,L,
\\
H_\ell = H_0 + \sum_{m=1}^Mc_{\ell m}(H_m\otimes I_{\nb}),\
\ell=1,\cdots,L
\eea
\eeq
where all the control amplitudes $c_{lm}=c_{m}(lT/L)$ are collected in
the vector $c\in\Rbf^{LM}$. Note that piecewise constant controls are
not crucial for the landscape analysis that follows. Any suitable
parameterization of the control fields $c_{m}(t)$ will have a similar
effect, \eg, via frequencies, amplitude and phases of a Fourier
series.

\section{Fidelity functions}
\label{sec:fid}

The design objective is to find control parameters $c\in\Rbf^{LM}$ so
that the final time unitary matrix $U(c)$ is as close as possible to a
desired unitary evolution $W\in \text{U}(N_{A})$ on system $A$.  If
this is achieved exactly then the final time unitary matrix will factorize,
\ie, $U(c)=W\otimes \Phi$ for some $\Phi\in \text{U}(N_{B})$.  The
goal of this paper is to find a condition for when the search for
controls achieving this goal is trap-free.

To quantify the search we proceed by introducing a fidelity function
for a bipartite quantum system and formulate the corresponding
optimization problem. The introduced fidelity function characterizes
the control landscape of a bipartite quantum system so that we can
define subsequently what we mean by a trap-free search. We distinguish
between a model-based and data-based control design procedure. We show
that there is a fidelity function variable that is common to both.

\subsection{Model-based control design}

If we assume that \refeq{uhc} is an accurate model of the bipartite
system, the optimization problem to find the controls $c$ to implement
$W$ on the $A$-system can be formulated as minimizing, with respect to
$(c,\Phi)\in\Rbf^{LM}\times\U(\nb)$, the distance $D$ between $U(c)$
given by \eqref{eq:uhc} and a decoupled target system $W\otimes
\Phi$. Specifically, we use the distance measure developed in
\cite{GDKBH:10},
\beq[eq:optDist]
D(c,\Phi) = \normfro{U(c)-\udes\otimes\Phi}^2
\eeq
where $\normfro{X}=\sqrt{\trace X^\dag X}$ is the Frobenius norm.  As
shown in \cite{GDKBH:10}, for a given control $c$, optimization over
$\Phi$ gives,
\beq[eq:optPhi]
\bea{rcl}
\ds
\min_{\Phi\in\U(\nb)}D(c,\Phi)
&=&
\ds
2N\left( 1-(1/N)\max_{\Phi\in\U(\nb)}J(c,\Phi) \right)
\\
&=&
\ds
2N\left( 1 - \sqrt{F(c)} \right)
\eea
\eeq
with,
\beq[eq:JcPhi]
\bea{l}
J(c,\Phi) = \real\trace\{(W\otimes\Phi)^\dag U(c)\}
\\
\ds
\max_{\Phi\in\U(\nb)}J(c,\Phi) = \normnuc{\Gam(c)}
\eea
\eeq
and
\beq[eq:Fnuc]
F(c) = \normnuc{\Gam(c)/N}^2,
\eeq
where $\normnuc{X}=\trace\sqrt{X^\dag X}$ is the nuclear norm (\ie,
the sum of the singular values of $X$), and where $\Gam(c) =
\sum_{a=1}^{\na} R(c)_{[aa]} \in\mathbb C^{\nb\times \nb}$, $R(c) =
(W\otimes I_{\nb})^\dag U(c)\in\U(\n)$ with
$\{R(c)_{[aa]},~a=1,\cdots, \na\}$ the $\na$ block diagonal
$\nb\times\nb$ submatrices of $R(c)$.

We refer to (i) $F(c)$ as a function of $c\in\Rbf^{LM}$ as the
\emph{control landscape}, (ii) $J(c,\Phi)$ as a function of
$(c,\Phi)\in\Rbf^{LM}\times\U(\nb)$ as the \emph{extended control
  landscape}, and (iii) $\Phi$ as the \emph{extended landscape unitary
  matrix}.  Both of these functions are bounded: $F(c)\in[0,1]$ and
$J(c,\Phi)\in[-N,N]$.

The unitary matrix $\Phiopt$ which achieves the above maximum is
obtained from the singular value decomposition $\Gam(c) = T_{\rm
  left}QT_{\rm right}^\dag$ with $T_{\rm left},T_{\rm right}\in\U(\n)$
and $Q = \diag(q_1,\ldots,q_{N_{B}})\geq 0$, so that we obtain,
\beq[eq:Phiopt]
\bea{rcl}
\Phiopt(c)= T_{\rm left}T_{\rm  right}^\dag.
\eea
\eeq 
For our subsequent analysis it is more convenient to parameterize the
unitary matrix $\Phi$ via the generator $\mathcal
B(\phi)=\sum_{b=1}^{N_{B}^{2}}\phi_{b}B_{b}$ by,
\beq[eq:Phib]
\Phi(\phi) = \exp\{i\Bcal(\phi)\},
\eeq
where $\{B_{b}\}_{b=1}^{N_{B}^{2}}$ is an operator basis for system
$B$.  The real parameters $\phi_{b}$ are collected in the vector
$\phi\in\mathbb R^{N_{B}^{2}}$ referred to as the \emph{extended
  landscape parameter}. The extended landscape objective is then
equivalent to,
\beq[eq:Jcphi]
J(c,\phi) = \real\trace\{\left(W\otimes\Phi(\phi)\right)^\dag U(c)\}.
\eeq
In each appropriate context we sometimes use ``fidelity'' when
referring to $F(c)$, $J(c,\Phi)$, or $J(c,\phi)$.

The maxima $F(c)=1$ and $J(c,\phi)=N$ (equivalently $D(c,\phi)=0$) are
obtained iff the target unitary matrix $W$ on system $A$ is achieved
for the final time unitary matrix, \ie, $U(c)=W\otimes U_B$ for some
unitary matrix $U_{B}\in\U(\nb)$. When this occurs we also have
$\Gam(c)=N_{A} U_B$ and $\Phiopt(c)=\Phi(\phiopt(c)) = U_B$.

For a \emph{closed} system, \ie, there is no $B$-system present,
$\nb=1$, $\n=\na$, and the extended landscape variable
$\Phi=e^{i\phi}$ is reduced to a global phase on the target unitary matrix
$W$. Consequently, $\Gam(c)=\trace\ W^\dag U(c)$ and $\max_\phi
J(c,e^{i\phi}) = |\trace\ W^\dag U(c)|$ which when squared and
normalized gives $F(c)=|\trace\ \{W^\dag U(c)\}/\na|^2\in[0,1]$, an
often used fidelity measure for closed quantum systems.


\subsection{Data-based control design}

If the model \refeq{uhc} is not known adequately, the control fields
can be directly learned in an experiment using data obtained from
measurement outcomes \cite{JudsonRabitz:92}. Assuming that the initial
state $\rho_{B}$ of system $B$ is uncorrelated with the initial state
of system $A$ we can use quantum process tomography (QPT) to estimate
the time evolution of system $A$, which is in general given by a
quantum channel that is a completely positive trace preserving map
\cite{nielsen2002quantum, poyatos1997complete} represented by a
process matrix $X(c)\in\mathbb C^{N_{A}^{2}\times N_{A}^{2}}$.  As
shown in \cite{GDKBH:10}, the \emph{channel fidelity} $\hat{F}(c)$,
measuring how close the quantum channel is to the target unitary $W$
on system $A$, is given by,
\beq[eq:Fdata]
\hat{F}(c) = (1/N_{A}^{2})\vec{W}^\dag X(c)\vec{W}
= (1/N_{A}^{2})\trace\{\Gam(c)\rhob\Gam(c)^\dag\},
\eeq  
where $\vec{W}\in\Cbf^{\nasqr}$ is the vectorized version of the
target unitary $W\in\U(\na)$.  Note that \emph{only} $X(c)$ is known
from the data, whereas $\Gam(c)$ and $\rhob$ are not. However, this
fidelity, like \refeq{Fnuc} is a direct measure of a norm of
$\Gam(c)$: in \refeq{Fnuc} the norm is the square of the sum of the
singular values, whereas in \refeq{Fdata} it is a weighted sum of the
singular values squared. In either situation these fidelities are in
the range $[0,1]$ and the maximum of 1 is only achieved when $W$ is
implemented on system $A$. Notwithstanding the intermediate step of
QPT, because \refeq{Fnuc} and \refeq{Fdata} effectively measure a norm
of $\Gam(c)$, their landscape properties are essentially the same,
modulo some minor effects on the critical points as discussed in
\cite{DominyETAL:2011}.

\section{Trap-free search}
\label{sec:trapfree}

Searching over the control parameters $c\in\Rbf^{LM}$ to maximize the
model-based fidelity $F(c)$ from \refeq{Fnuc} or the data-based
fidelity $\hat{F}(c)$ from \refeq{Fdata} is in general not a convex
optimization problem due to the inherent bilinear control structure of
the Hamiltonian \eqref{eq:totalH}.  There are many approaches
\cite{Brif.NJP.12.075008.2010} to determining $c$ including stochastic
algorithms \cite{JudsonRabitz:92} and gradient-based algorithms
\cite{khaneja2005optimal,KosutGB:13}, \etc~ In almost all cases the
search for controls with various objectives is proving to be
remarkably successful, both in simulations and in a variety of
experiments. As a result, no matter how the search for the controls is
conducted, we would like to know under what conditions the landscape
is trap-free, \ie, when there are no sub-optimal local
maxima. Specifically, by a \emph{trap-free} search we mean the
following:
\bquote \emph{\noindent The control landscape $F(c)$ is trap-free if
  the critical points (where $\nabla_cF(c)=0$) are either saddles or
  global extrema.}  \equote
Towards reaching this latter assessment, we defined two search
landscapes: the control landscape with objective $F(c)$ from
\refeq{Fnuc}, and the extended landscape with objective $J(c,\phi)$
from \refeq{Jcphi}.  We will show that at the optimal parameter
$\phiopt(c)$ (obtained from \refeq{Phiopt} via
$\Phiopt(c)=\exp\{i\Bcal(\phiopt(c))\}$), when the extended landscape
$J(c,\phiopt(c))$ is trap-free, then so is the control landscape
$F(c)$.

\section{Critical Points of the Control Landscape}

To see how a trap-free search of the control landscape \refeq{Fnuc}
could arise, we first examine in more detail the character of the
critical points of the extended landscape fidelity \refeq{Jcphi}.

\subsection{Spectral decomposition}

The extended landscape $J(c,\phi)$ from \refeq{Jcphi} depends on a
unitary matrix denoted by $\Uobj(c,\phi)\in\U(\n)$ which can be
decomposed via,
\beq[eq:vomv]
\bea{rcl}
\Uobj(c,\phi)
&=&
(W\otimes\Phi(\phi))^\dag U(c)
\\
&=& V(c,\phi)e^{i\Om(c,\phi)}V(c,\phi)^\dag,
\eea
\eeq
where $\Om(c,\phi)=\diag(\om_1(c,\phi),\cdots,\om_N(c,\phi))$, the
unitary matrix satisfies $V(c,\phi)\in\text{U}(N)$, and spectral
frequencies $\{\omega_{n}(c,\phi)\}$ are conveniently collected in the
vector $\om(c,\phi)\in\Rbf^N$. The fidelity then simply becomes,
\beq[eq:Jcos]
\bea{rcl}
J(c,\phi)=\sum_{n=1}^{N}\cos(\omega_{n}(c,\phi)). 
\eea
\eeq
and the gradient of $J(c,\phi)$ takes the form
\beq[eq:dJcphi]
\bea{rcl}
\nabla_{c,\phi}J(c,\phi)
&=&
G_{c,\phi}(c,\phi)g(\om(c,\phi))\in\Rbf^{LM+\nbsqr}, 
\eea
\eeq
where, following the terminology in \cite{RabitzHR:04}, the matrix
\beq[eq:Gcphi]
G_{c,\phi}(c,\phi) = \nabla_{c,\phi}\om(c,\phi) =
\left[\bea{c}G_c(c,\phi)\\G_\phi(c,\phi)\eea\right]
\in\Rbf^{(LM+N_B^2)\times N}
\eeq
is referred to as the \emph{dynamic gradient} and the vector
\beq[eq:gom]
g(\om)=
-(\sin(\omega_{1}(c,\phi)),\cdots,\sin(\omega_N{}(c,\phi)))^{T}\in\mathbb R^{N}
\eeq
as the \emph{kinematic gradient}.  Detailed expressions for the
constituant dynamic gradient matrices
$G_c(c,\phi)=\nabla_c\om(c,\phi)\in\Rbf^{LM\times N}$ and
$G_\phi(c,\phi)=\nabla_\phi\om(c,\phi)\in\Rbf^{N_B^2\times N}$ are
presented in Appendix \refsec{dyngrad} along with some of their
properties .

\subsection{Kinematic critical points}
\label{sec:kincrit pts}

In terms of the vector of spectral frequencies $\omega\in\Rbf^\n$ we
define the \emph{kinematic fidelity} in \refeq{Jcos} by
$J(\om)=\sum_n\cos\om_n$, and the associated \emph{kinematic gradient}
by $\nabla_\om J(\om)=g(\omega)$, so that the \emph{kinematic Hessian}
is given by $\nabla^2_\om
J(\om)=-\diag(\cos(\omega_{1}(c,\phi)),\cdots,\cos(\omega_{N}(c,\phi)))$. A
kinematic critical point at which $g(\omega)=0$ is obtained iff
$\sin(\omega_{n}(c,\phi))=0$ for all $n$ (equivalently
$\cos(\omega_{n}(c,\phi))=\pm 1$).  Using standard trigonometry
(Appendix \refsec{kinJ}) yields the following picture. Trivially the
maximum and minimum of $J$ is given by $J(\omega)=\pm N$ with a
corresponding Hessian $\nabla_{\omega}^{2}J(\omega)=\mp I_{N}$,
respectively. The interior is defined through $J(\omega)=(2p-N)$ and
$\nabla_{\omega}^{2}J(\omega)=\diag(-I_{p},I_{N-p})$ for $1\leq p\leq
N-1$. Thus, the kinematic critical points associated at the top and
bottom of the landscape where $J(\om)=\pm N$ are global extrema
because the Hessians there are, respectively, negative and positive
definite. In the interior of the landscape where $J(\om)\in(-N,N)$
there are no traps, only saddles; there the Hessians are all
indefinite with both positive and negative eigenvalues of unit
magnitude. Moreover, these critical points produce exactly $N-1$
critical \emph{values} at which $J(\om)\in\{2-N,\ldots,N-2\}$.
However, there may be an infinite number of frequencies and/or control
combinations which produce the $N-1$ saddle point objective values.
Landing on one of these saddles is highly unlikely with the gradient
algorithm, so these are usually of no practical importance; however
their presence can influence the efficiency of climbing the landscape.

\section{Rank conditions for a trap-free search}
\label{sec:ranktrapfree}

To arrive at our main result first note that from
\refeq{JcPhi}-\refeq{Fnuc} together with \refeq{Phib} where
$\Phiopt(c)=\Phi(\phiopt(c))$, the extended landscape objective at
$\phiopt(c)$ becomes,
\beq[eq:Jopt]
\bea{rcl}
J(c,\phiopt(c)) &=& \normnuc{\Gam(c)}=\n\sqrt{F(c)},
\eea
\eeq
and the gradient with respect to the controls $c$ then reads,
\beq[eq:dJcopt]
\bea{rcl}
\nabla_cJ(c,\phiopt(c)) &=& \left(\n/\sqrt{F(c)}\right)\nabla_cF(c).
\eea
\eeq
This result shows that at the optimal $\phiopt(c)$ the extended
landscape fidelity is positive, and additionally (see Appendix
\refsec{dJphi}) we also find that the following two landscape
properties must hold:
\beq[eq:atphiopt]
\bea{l}
\sum_{n=1}^\n\sin\om_n(c,\phi_\opt(c))=0,
\\{}\\
\nabla_\phi J(c,\phi_\opt(c))
= G_\phi(c,\phi_\opt(c))g(\om(c,\phi_\opt(c))
=0,
\eea
\eeq
The second property above follows by definition of $\phiopt(c)$: the
extended landscape objective is maximized by $\max_\phi J(c,\phi) =
J(c,\phi_\opt(c))$. Consequently, at $\phiopt(c)$, the gradient of the
extended landscape objective with respect to the extended landscape
parameters is zero. Together with the definitions of the gradient
matrices in \refeq{dJcphi}, the equivalent expression for the
gradient in \refeq{Jopt} is,
\beq[eq:dJcphiopt]
\nabla_cJ(c,\phiopt(c)) = G_c(c,\phiopt(c))g(\om(c,\phiopt(c)),
\eeq
so that with \refeq{Gcphi} and \refeq{dJcopt}-\refeq{atphiopt} we
arrive at,
\beq[eq:dJcphiGopt]
\bea{rcl}
\left.\nabla_{c,\phi}J(c,\phi)\right|_{\phi=\phiopt(c)}
&=&
\left[
  \left.\bea{c}G_c(c,\phi)\\G_\phi(c,\phi)\eea\right]
g(c,\phi)\right|_{\phi=\phiopt(c)}
\\&&\\
&=&
\left[
  \bea{c} \left(N/\sqrt{F(c)}\right)\nabla_cF(c)
  \\\bold{0}\eea
  \right]
\eea
\eeq
Since the extended landscape gradient at the optimal landscape
parameter as expressed above is a product of the dynamic and the
kinematic gradient \refeq{dJcphi}-\refeq{gom}, and since the kinematic
gradient $g(\om)$ is zero \emph{only} at the global extrema $J=\pm\n$
or at saddles, thereby forming a \emph{sufficient condition} to ensure
that there are no traps (local extrema) in the control landscape
interior $F(c)\in(0,1)$ is reflected in the following rank condition
for the dynamic gradient $G_{c,\phi}(c,\phiopt(c))$.

\vspace{1ex}
\noindent
\textbf{Rank condition for trap-free search} \emph{Assume that the
  dynamic gradient of the extended landscape, $G_{c,\phi}(c,\phi)$
  evaluated at the optimal extended landscape parameter satisfies,
\beq[eq:rankGcphiopt]
\bea{lll}
\rank G_{c,\phi}(c,\phiopt(c))
&=N, & \Uobj(c,\phiopt(c))\in{\rm U}(N)
\\
&\geq N-1, & \Uobj(c,\phiopt(c))\in\SU(N)
\eea
\eeq  
with $\Uobj(c,\phi)$ from \refeq{vomv}. Under this condition, the
control landscape $F(c)$ is trap-free, meaning that $\nabla_cF(c)=0$
either at a saddle where $F(c)\in(0,1)$ or at the global
extrema. including the global maximum $F(c)=1$.}
\vspace{3ex}

\noindent
We emphasize again that the rank condition \refeq{rankGcphiopt} is a
sufficient condition for a trap-free search. We have not established
the conditions which guaranty that \refeq{rankGcphiopt} holds.  Moreover,
it is a strong condition in the sense that it does subsume any prior
assumptions about controllability, reachability, and control
resources. To emphasize this point, if, in fact, \refeq{rankGcphiopt}
holds, then a gradient algorithm \emph{will} reach the top of the
control landscape; \emph{a fortiori}, a final-time decoupling control
exists.

Interestingly, if \refeq{rankGcphiopt} holds, then in the landscape
interior $F(c)\in(0,1)$, except at saddles, the fidelity gradient
satisfies $\nabla_c F(c)\neq 0$ even if the rank of the dynamic
control gradient $G_c(c,\phi_\opt(c))$ is not full.  Recall that the
matrix $G_{c}$ describes the gradient of the unitary frequencies
$\omega(c,\phi)$ with respect to $c$ with $\phi$ fixed, where the
controls only act on system $A$. Intuitively we expect that when the
total $AB$ system cannot be \emph{fully} controlled by acting on $A$
alone, the frequencies cannot arbitrarily be varied as a function of
$c$. Furthermore, we expect that then $G_{c}$ is not full
rank. However, the rank condition \refeq{rankGcphiopt} can hold by
virtue of the apparent \emph{additional flexibility} offered by
$G_\phi(c,\phi_\opt(c))$, the dynamic gradient with respect to the
extended landscape parameter. Clearly the $\nbsqr$ extended landscape
parameters can offset rank deficiencies due to the $LM$ control
parameters alone.

At the top we have: $F(c)=1$, the final time unitary $U(c)=W\otimes
U_B$, the optimal extended landscape variable $\Phiopt(c)=U_B$, and
the extended landscape unitary $\Uobj(c,\phiopt(c)) = I_N$. This means
that as a search converges to the top of the landscape,
$\Uobj(c,\phiopt(c)) \to I_N$ (clearly in $\SU(N)$), and so the sum of
the spectral frequencies converges to zero, and hence, the rank
condition converges to $\rank~G_{c,\phi}(c,\phiopt(c))\geq N-1$.

An interesting special case of $\Uobj\in\SU(N)$ is when the spectral
frequencies are symmetric about zero and otherwise unequal. When this
occurs the rank condition for a trap-free search drops to $\n/2$. To
see how this comes about, assume that the spectral frequencies are
ordered so that $\om=[-\omb,\ \omb]$ with $\omb_1 > \omb_2 > \cdots >
\omb_{\n/2}>0$. It follows that the dynamic gradient has the form
$G(c,\phi)=[-\bar{G}(c,\phi)\ \bar{G}(c,\phi)]$ with
$\bar{G}(c,\phi)\in\Rbf^{(LM+\nbsqr)\times \n/2}$. Hence, under these
condition the control landscape is trap-free if $\rank
G_{c,\phi}(c,\phi_\opt(c))\geq \n/2$.  We show an example of this
reduced rank condition in \figref{identity} of a single spin coupled
to a random bath with the identity target.


Before presenting numerical support for the main result, it is
informative to examine \refeq{rankGcphiopt} for a (non-bipartite)
closed-system. Since there is no $B$ system we have $\nb=1$, $\n=\na$,
and $\Phi(\phi)$ reduces to a global phase on the target unitary
matrix, \ie, the extended landscape unitary matrix becomes
$\Uobj(c,\phi)=e^{i\phi}W^{\dag} U(c)$. As shown in Appendix
\refsec{closed sys} the dynamic gradient simplifies to,
\beq[eq:Gcph closed]
G_{c,\phi} = \left[\bea{c}G_c\\
  \bold{1}
  \eea\right]\in\Rbf^{(LM+1)\times\n}
\eeq
from which is obtained,
\beq[eq:rankGcphi closed]
\rank G_{c,\phi} = \min\{\rank G_c + 1,\n\}.
\eeq
Thus, if for a closed quantum system $\rank G_c\geq N-1$ is satisfied,
a gradient algorithm will find the controls that implement a target
unitary transformation up to global phase. We emphasize here again
that no statement is made about the validity of this assumption on the
rank of $G_c$. However, assuming that sufficient control resources are
available, it has recently been shown that almost all closed quantum
systems are trap free \cite{RussellRW:17}. That is, if one picks the
drift and the control Hamiltonian at random, and furthermore, there
are no control field constraints, the control landscape of almost all
(except a null set) closed systems is trap free. We therefore expect
that when sufficient control resources are available, the condition
$\rank G_c \geq N-1$ holds for almost all closed quantum systems,
though a rigorous proof connecting the results in \cite{RussellRW:17}
and the analysis presented above for closed systems is left for future
studies.
 
The single extra degree of freedom due to the extended landscape
parameter $\phi$ makes it clear that $G_c$ can have a reduced
rank. For example, if $\Uobj\in\SU(N)$, the landscape is trap-free
even if the rank of $G_c$ is as as low as $N-2$, provided that $N-2
\leq LM$.  The numerical results to follow suggest that this
flexibility is more pronounced for bipartite system.

\section{Numerical examples}

In order to find the controls that implement a desired target unitary
on system $A$ we use a gradient algorithm starting with an initial
control $c^{0}\in\Rbf^{LM}$ which we choose throughout this section to
be the zero vector, and iterate for $i=1,2,\ldots$, according to,
\beq[eq:gga]
\bea{rcl}
c^{i} &=& c^{i-1} + \gam_i\nabla_cF(c^{i-1}),
\eea
\eeq
where the gradient is obtained numerically. The step-size, $\gam^i$,
is increased, and the control update accepted, whenever $F(c^i) >
F(c^{i-1})$. Otherwise the control update is not accepted, the
step-size is decreased and \refeq{gga} is repeated with the previous
control. The algorithm is halted when $F(c^i)$ is insufficiently
increasing. In order to see whether the rank condition holds, the
singular values of $G_{c,\phi}$ and $G_{c}$ at $\phi_{\text{opt}}$ are
calculated during the iteration.

\subsection{Central spin system}
\label{sec:cspin}

We begin by investigating a single spin ($N_{A}=2$) that interacts
through a Heisenberg type interaction with $q_B$ environmental spins
$(N_{B}=2^{q_{B}})$ and a single control $c(t)$ is applied in the $z$
direction on the system spin. The total Hamiltonian describing the
control system reads
\beq[eq:cspin]
\bea{rcl}
H(t) &=& \sig_y\otimes I_{N_{B}} + \sum_{q=1}^{q_B}
a_q\sum_{s=x,y,z}\sig_s\otimes\sig_s^{(q)}
\\
&&
+ c(t)\left(\sig_z\otimes I_{N_{B}}\right),
\eea
\eeq
where $\sigma_{s},~s\in\{x,y,z\}$ are the Pauli spin operators.  The
controllability aspects of this model, also called the \emph{central
  spin model}, were studied in \cite{ArenzGB:14}. It was shown that
for all values of the coupling constants $a_q$, and independent of the
number $q_B$ of environmental spins, the system spin (i.e., the system $A$) is
fully controllable. Moreover, if all the coupling constants are
different from each other, the total system (i.e., the system spin +
environmental spins) becomes fully controllable.

\begin{figure}[h]
 \includegraphics[width=0.8\columnwidth]{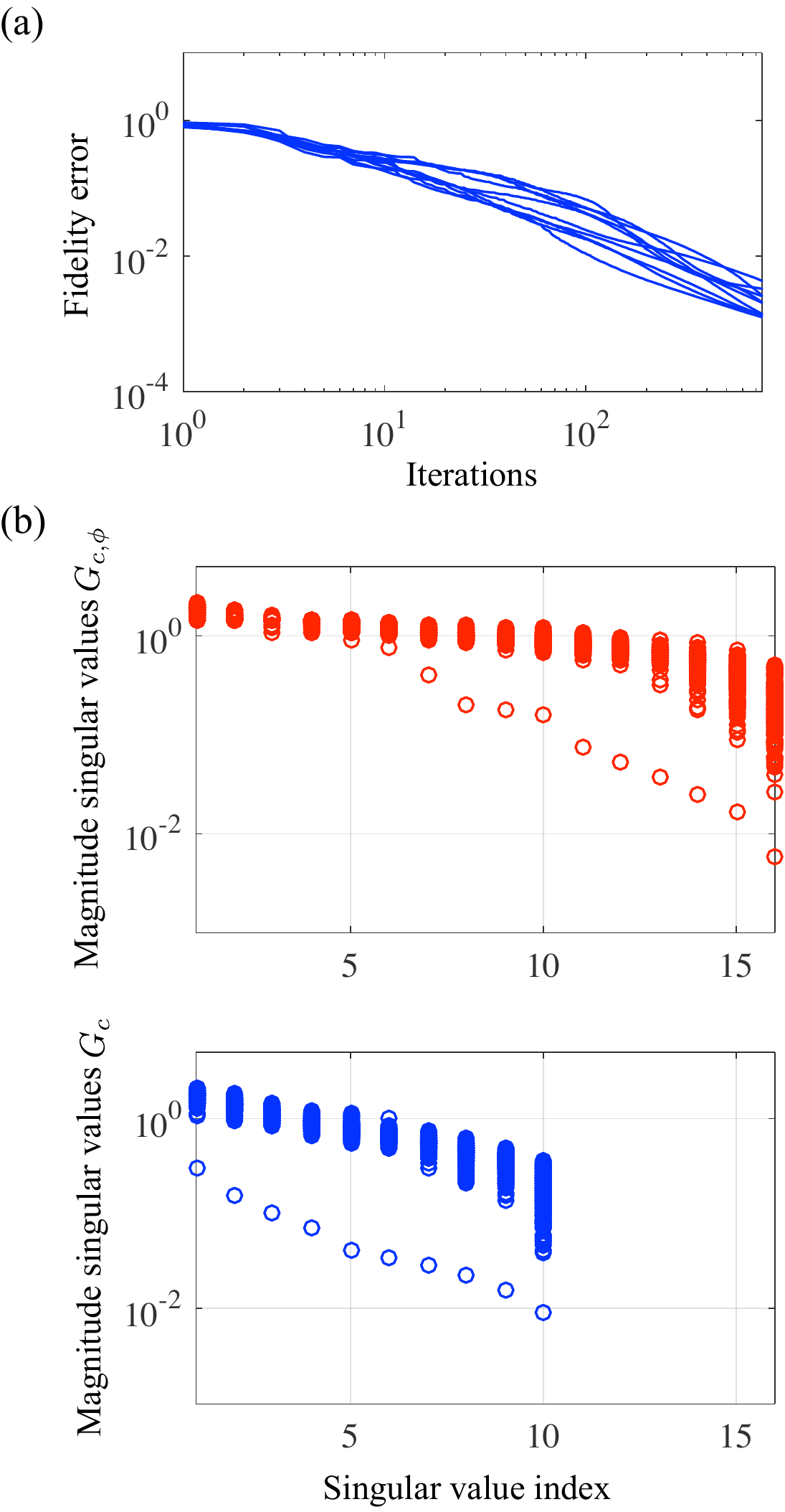}
 \caption{\label{fig:centralspin} Numerical investigation of the
   convergence of the gradient algorithm \eqref{eq:gga} for the
   central spin system \eqref{eq:cspin} with $q_{B}=3$ and equal
   coupling constants $a_{q}$. (a) Fidelity error $1-F$ as a function
   of the iteration step for 10 randomly chosen unitary
   transformations as a target on the system spin, for a fixed
   evolution time $T=20$ and $L=100$ piecewise constant controls. (b)
   Singular values (circles) of the gradient $G_{c,\phi}$ and the
   gradient $G_{c}$ at $\phi_{\text{opt}}$ during the iteration. The
   vertical spread of the values correspond to singular values
   during the iteration process for the different targets. The number
   of singular values of $G_{c,\phi}$ is seen to be fully consistent
   with the assumption \refeq{rankGcphiopt}, \ie,
   $\rank~G_{c,\phi}=N=16$ even though $\rank~G_c=10$.}
\end{figure}

In Fig. \ref{fig:centralspin} we studied the implementation of some
randomly chosen unitary transformation $W\in \text{U}(2)$ on the
system spin for a total evolution time of $T=20$ with $L=100$
piecewise constant controls and the coupling constants $a_{q}$ in
\eqref{eq:cspin} all chosen to be equal. We remark here that, even
though the control system \eqref{eq:cspin} evolves on $\text{SU}(N)$,
the ability to additionally maximize over $\phi$ allows for
implementing $W$.  Fig. \ref{fig:centralspin} (a) shows the fidelity
error $1-F$ as a function of the iteration step $i$, whereas each
curve corresponds to a randomly chosen target unitary. The results
suggest that independently of the target transformation on the system
spin, the gradient algorithm finds the controls that implement the
target unitary (up to some error which continues to decrease when the
iterations stopped), even though the total system is not fully
controllable. We remark here that the same behavior was found for
additional numerical simulations using different $q_{B}$ and $a_{q}$,
which suggest that traps rarely occur in the control landscape for the
central spin model. We now turn to whether this behavior is in
agreement with the sufficient condition \eqref{eq:rankGcphiopt} for a
trap free search. In Fig. \ref{fig:centralspin} (b) we show the
singular values (circles) of $G_{c,\phi}$ and $G_{c}$ at
$\phi_{\text{opt}}$ determined by \eqref{eq:Phiopt} during the
iterative process. We see that the singular values displayed are
consistent with the corresponding rank condition expressed in
\refeq{rankGcphiopt}, \ie, for $N=16$, $\rank~G_{c,\phi}=16$ despite
$\rank~G_c =10$. As we have previously remarked, it is unlikely
  that the total $AB$ system can be \emph{fully} controlled by acting
  on $A$ alone; the frequencies cannot arbitrarily be varied as a
  function of $c$.

\subsection{Single qubit: random bath with identity gate}
\label{sec:1quxz}

In the last subsection we investigated the central spin model for
which the system $A$ spin is fully controllable with a single control
field. However, as mentioned before, the rank condition
\eqref{eq:rankGcphiopt} for a trap free search has no evident direct relation
to whether the system $A$ is fully controllable or not. Now we
consider a system $A$ which is not fully controllable with a single
control field. That is, we consider a spin interacting with a random
bath of dimension $N_{B}=8$ described by the total Hamiltonian
\beq[eq:quxz]
H(t) = c(t)\sig_x\otimes I_{N_{B}} + \sig_z\otimes B_z,
\eeq
where $B_{z}$ with $\Vert B_{z}\Vert=1$ is a randomly chosen Hermitian
matrix. We remark here that such types of Hamiltonians are typically
used to model pure dephasing of a single spin interacting with a
spin-bath. For instance, in nitrogen vacancy centers dephasing of the
electron spin caused by an interaction with surrounding nuclear spins
can be modeled by an interaction Hamiltonian of the form
$H=\sigma_{z}\otimes\sum_{q}a_{q}\sigma_{z}^{(q)}$. Even though the
system spin is not fully controllable with a single control field
$c(t)$, which can easily be seen from the underlying dynamical Lie
algebra \cite{elliott2009bilinear, d2007introduction}, at any time $T$
the identity operation can be implemented on the system spin, for
instance, by using delta function like controls that invert the sign
of the interaction part. Even though the identity operation on $A$ is
a very specific choice for a target operation, such a target operation
is of practical importance in the context of dynamical decoupling,
\ie, suppressing the interactions with system $B$.

\begin{figure}[h]
 \includegraphics[width=0.8\columnwidth]{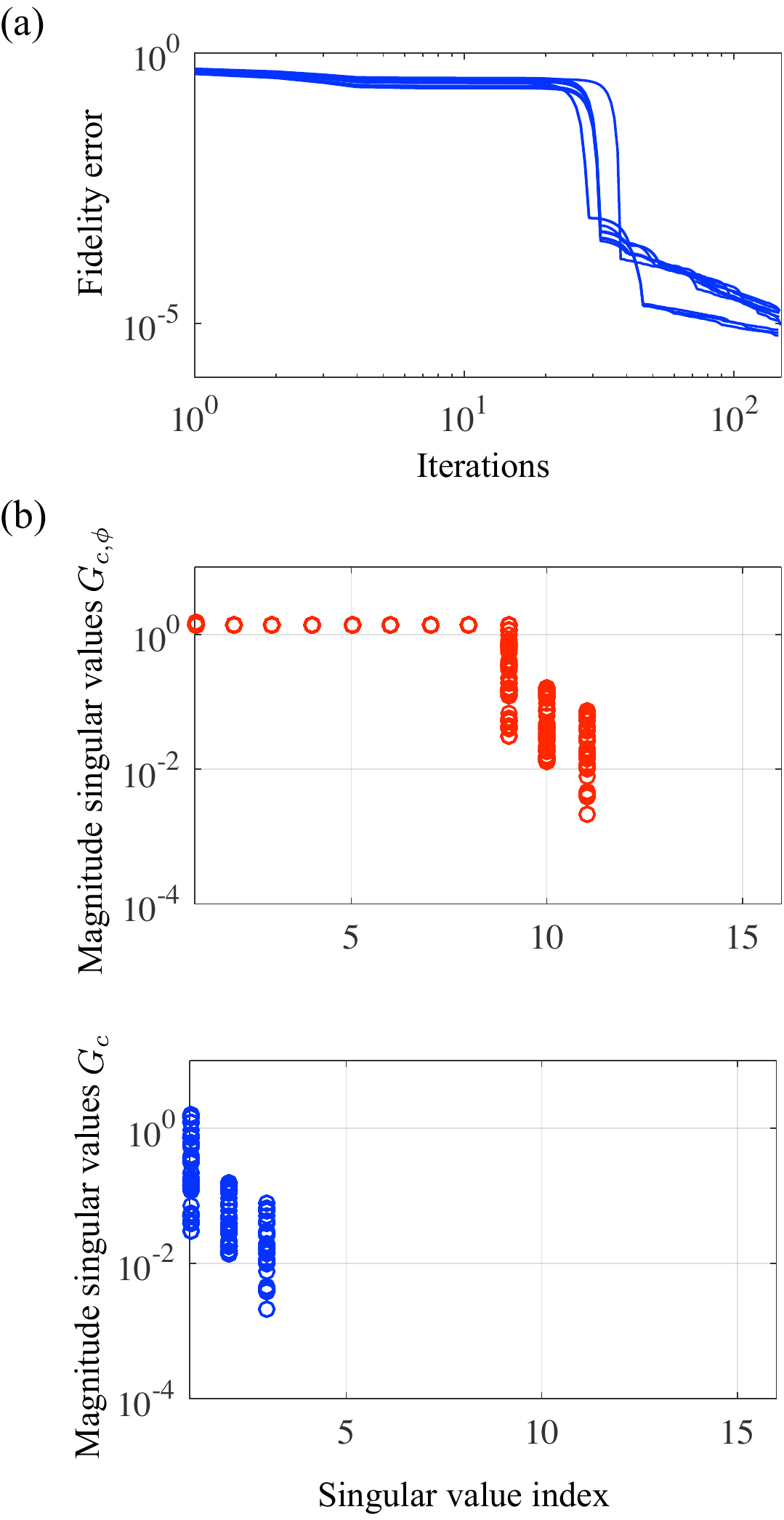}
 \caption{\label{fig:identity} Numerical investigation of the
   convergence of the gradient algorithm for the random bath model
   \eqref{eq:quxz} where $B_{z}$ with $\Vert B_{z}\Vert=1$ is randomly
   chosen Hermitian $8\times 8$ matrix. (a) Fidelity error as a
   function of the iteration step for 10 randomly chosen $B_{z}$'s,
   for a fixed evolution time $T=1$ and $L=4$ number of piecewise
   constant controls. (b) Singular values (circles) of the gradient
   $G_{c,\phi}$ and the gradient $G_{c}$ at $\phi_{\text{opt}}$ during
   the iteration. The vertical spread of the values correspond to
   singular values during the iteration process for the different
   targets. The number of singular values of $G_{c,\phi}$ is seen to
   be fully consistent with the assumption \refeq{rankGcphiopt}, \ie,
   $\rank~G_{c,\phi}=11\geq N/2=8$ even though $\rank~G_c=3$.}
\end{figure}

In Fig. \ref{fig:identity} we show results for the implementation of
the identity operation on the system spin with a total evolution time
$T=1$ and $L=4$ piecewise constant controls with $B_z$ a randomly
chosen $\nb\times\nb$ Hermitian matrix normalized to $\norm{B_z}=1$
and with $\nb=8$. Fig. \ref{fig:identity} (a) shows the fidelity error
$1-F$ as a function of the iteration step $i$, where each curve
corresponds to a randomly chosen $B_{z}$. The curves show that for all
choices of $B_{z}$ the gradient algorithm finds the controls that
implement the identity. In Fig. \ref{fig:identity} (b) we show the
singular values (circles) of $G_{c,\phi}$ and $G_{c}$ at
$\phi_{\text{opt}}$ during the iteration process. Because we chose the
target to be the identity and the control system \eqref{eq:quxz} is
expressed though Pauli operators, the spectral frequencies $\omega$
are symmetric around zero. As previously noted, when this occurs the
rank condition for a trap-free search drops to $N/2=8$. From
Fig. \ref{fig:centralspin} (b) we see that this condition is
satisfied, \ie, with $N=16$, $\rank~G_{c,\phi}=11 > 8$ despite
$\rank~G_c =3$.

\section{Summary and Outlook}

We have investigated the quantum control landscape of a bipartite
quantum system where only one part, $A$, is accessible through time
dependent controls; part $B$ is not. The control objective was to
implement a desired unitary operation on part $A$ of the bipartite
system, regardless of which unitary transformation is implemented on
part $B$. By defining an extended control landscape, and along with
the controls, additionally optimizing over the corresponding extended
landscape parameters, we found that if a sufficient condition is
satisfied for the extended landscape to be trap-free, then it implies
that the original landscape is also trap-free. Specifically, since the
kinematic gradient only possess saddles in the interior of the
landscape, if the rank of the dynamic gradient at the optimal extended
landscape parameters is full, in the particular sense specified, a
gradient search on the original landscape will converge to the global
maximum ($F=1$). In addition, it is not necessary for the control
gradient $G_{c}$ to have full rank. As we have noted, the developed
rank condition on the extended landscape can hold by virtue of the
apparent flexibility offered by maximizing over the parameters that
define the extended landscape.  We have provided two common supporting
numerical examples for which the rank condition holds. Although we
have restricted the presentation to a bipartite $AB$-system with
\emph{both} control and target on the $A$-system, other system
combinations are clearly possible. For example, if the controlled
system space consisted of target and ancilla qubits, the latter to
help facilitate error reduction, then the extended landscape unitary
$\Phi$ would have the dimension of the ancilla-bath system.

As we have remarked in various places throughout the text, the rank
condition \refeq{rankGcphiopt}, presented as an assumption, is
sufficient upon satisfaction to ensure a successful gradient search to
the top of the landscape. Though it can be used informally to test
various control sequences and/or search algorithms, establishing
properties of specific physical implementations, \ie, Hamiltonians,
for which \refeq{rankGcphiopt} is valid is an open research issue.  In
contrast to the closed system case, we do not expect an almost
``always statement'' \cite{RussellRW:17} to hold for bipartite control
systems. If the rank condition would hold for almost all bipartite
systems with controls applied on system $A$ only, this would imply
that every unitary transformation can almost always be perfectly
implemented on $A$ (full controllability), regardless of the form of
the control Hamiltonians as well as the interactions with system
$B$. Answering the question when system $A$ is fully controllable in
the presence of a $B$ system is still an open research matter.  Even
though we do not expect that an almost always statement exists for
bipartite systems, the detailed expressions of the dynamic gradients
given in the Appendix may provide a path forward for identifying
physical conditions under which the developed rank condition holds.
  
\textbf{Acknowledgments} The authors thank Benjamin Russell for many
helpful discussions and comments during the drafting of this
work. C. A. was supported by the NSF grant CHE-1763198 and H. R. was
supported by the DOE grant DE-FG02-02ER15344.

\bibliographystyle{plain} 
\bibliographystyle{unsrt}
\bibliography{rlk}

\noindent 
\begin{appendix}
\section{Dynamic gradients}
\label{sec:dyngrad}

Detailed expressions for the dynamic gradients in \refeq{dJcphi} are
herein derived. We first present the expressions followed by the
derivation and mention some properties along the way.

\subsection{Expressions of dynamic gradients}

Dropping the $(c,\phi)$-dependencies in \refeq{dJcphi} for ease of
reading, the dynamic gradients can be expressed as follows:
\beq[eq:grad prod ca]
\bea{rcl}
G_{c,\phi} &=& \left[\bea{c}G_c\\G_\phi\eea\right]
\in\Rbf^{(LM+\nbsqr)\times \n}
\\
G_c &=& \nabla_c\om
= -(I_L\otimes\Hcal)~\Ccal~\Vcala \in\Rbf^{LM\times \n}
\\
G_\phi &=& \nabla_\phi\om = -\Pcal\Vcalb \in\Rbf^{\nbsqr\times \n}
\eea
\eeq
%
The matrices in $G_c$ are:
\beq[eq:cala]
\bea{l}
\Hcal  =
  \left[
  \bea{c}
  {\overrightarrow{H_1\otimes \Ib}}^\dag
  \\\vdots\\
  {\overrightarrow{H_M\otimes \Ib}}^\dag
  \eea
  \right]\in\Cbf^{M\times N^2}
  \\
  \Ccal = \diag\{\Ccal_1,\ldots,\Ccal_L\}\in\Cbf^{LN^2\times LN^2}
  \\
  \Ccal_\ell = \ds
\int_0^\del e^{i\tau H_\ell^T} \otimes e^{-i\tau H_\ell}d\tau
\in\Cbf^{N^2\times N^2}
\\
\Vcala = \left[
  \bea{c}
  \Vcal_{A,1}
  \\
  \vdots
  \\
  \Vcal_{A,L}
  \eea
  \right] \in\Cbf^{LN^2\times N}
\\
\Vcal_{A,\ell} = \left[
    v_{\ell 1}^*\otimes v_{\ell 1}^{}\
  \cdots\
  v_{\ell N}^*\otimes v_{\ell N}^{}
  \right]\in\Cbf^{N^2\times N}
\\
\left[v_{\ell 1} \cdots v_{\ell N}\right] = U_{\ell+1}\cdots U_L V
\in\U(\n^2)
  \eea
  \eeq
Note that $\Vcal_{A,\ell}^\dag\Vcal_{A,\ell}=I_N$, hence
$\Vcala^\dag\Vcala=LI_N$ and thus both $\Vcal_{A,\ell}$ and $\Vcala$
have rank $N$.
  
Likewise the matrices in $G_\phi$ are:
%
%
\beq[eq:Pcal]
\bea{l}
\Pcal =
  \left[
  \bea{c}
  {\overrightarrow{\Is\otimes P_1}}^\dag
  \\\vdots\\
  {\overrightarrow{\Is\otimes P_{\nbsqr}}}^\dag
  \eea
  \right] \in \Cbf^{\nbsqr\times N^2}
  \\
  P_b = \ds \int_{\tau=0}^1 e^{i\tau \Bcal(\phi)} B_b e^{-i\tau\Bcal(\phi)}d\tau
  \in\Cbf^{\nb\times\nb}
  \\
  \Vcalb = \left[v_1^*\otimes v_1 \cdots v_N^*\otimes v_N\right]
  \in\Cbf^{N^2\times N}
\eea
\eeq
with $v_n\in\Cbf^N$ a column of $V$ in \refeq{vomv}. Thus
$\Vcalb^\dag\Vcalb=I_N$ from which it follows that $\rank\Vcalb=N$.
In addition, 
\beq[eq:Pb]
\bea{rcl}
\vec{P}_b &=& \Kcal\vec{B}_b
\\
\Kcal &=& \int_0^1 e^{-i\tau \Bcal(\phi)^T}\otimes e^{i\tau \Bcal(\phi)}d\tau
\in\Cbf^{\nbsqr\times\nbsqr}
\eea
\eeq 
Conditions derived in \cite{RussellVR:18} show that $\Ccal$ is almost
always invertible and similarly for $\Kcal$.

\subsection{Derivation of gradient expressions}
The variables $(c,\phi)$ are to
be selected to maximize the extended landscape objective function,
\beq[eq:obj]
J(c,\phi) = \real\trace(W\otimes\Phi(\phi))^\dag U(c)
\eeq
The gradient of $U(c)$ in \refeq{uhc} with respect to each element
$c_{\ell m}$ of $c$ is (for ease of reading we drop the
$(c,\phi)$-dependence),
\beq[eq:dUdx]
\bea{rcl}
\nabla_{c_{\ell m}}U &=& -iU_1 \cdots U_\ell Q_{\ell m} U_{\ell+1}\cdots U_L 
\\
Q_{\ell m} &=& \int_0^{\del} e^{itH_\ell} 
(H_{m}\otimes\Ib) e^{-itH_\ell} dt
\eea
\eeq
Using the unitary decomposition \refeq{vomv} write
$(W\otimes\Phi)^\dag U = Ve^{i\Om}V^\dag$ together with \refeq{obj}
gives,
\beq[eq:objprod]
\bea{rcl}
J &=& 
\real\trace\ 
e^{i\Om} = \sum_{n=1}^N\cos\om_n
\\ 
\nabla_{c_{\ell m}} J
&=&
\imag\trace\ (W\otimes\Phi)^\dag U_1\cdots U_\ell Q_{\ell k}U_{\ell+1}\cdots U_L
\\
&=&
\imag\trace~V_\ell^\dag Q_{\ell m}V_\ell~e^{i\Om}
\\ 
V_\ell &=& \left\{
\bea{l}
U_{\ell+1}\cdots U_L V,\ \ell=1,\ldots,L-1
\\
V,\ \ell=L
\eea
\right.
\eea
\eeq
The last lines follow from $(W\otimes\Phi)^\dag=Ve^{i\Om}V^\dag
U^\dag$. Since $V_\ell^\dag Q_{\ell m}V_\ell$ is Hermitian and $\Om$
is real and diagonal, it follows that the diagonal elements
$(V_\ell^\dag Q_{\ell m}V_\ell)_{nn},n=1\:N$ are also real. Hence,
\beq[eq:gradobj om]
\nabla_{c_{\ell m}} J
=
\sum_{n=1\:N}\left(V_\ell^\dag Q_{\ell m}V_\ell\right)_{nn}\sin\om_n
\eeq
Comparing the above expression for the gradient with $\nabla_{c_{\ell
    m}} J=\nabla_{c_{\ell m}}\sum_n\cos\om_n =-\sum_n(\nabla_{c_{\ell
    m}}\om_n)\sin\om_n$ gives the elements of the gradient matrix
$G_c\in\Rbf^{LM\times N}$ as,
\beq[eq:gradom]
(G_c)_{\ell m,n} = 
-\left(V_\ell^\dag Q_{\ell m}V_\ell\right)_{nn}
= - v_{\ell n}^\dag Q_{\ell m}v_{\ell n}^{}
\eeq
where $v_{\ell n}\in\Cbf^N$ are columns of the unitary $V_\ell$ in
\refeq{objprod}, that is, $V_\ell=[v_{\ell 1} \cdots v_{\ell N}]$.
Using repeated applications of the ``vec'' operator $\vec{(\cdot)}$,
which stacks the columns of a matrix into a vector, we arrive at the 
expression for $G_c$ is \refeq{grad prod ca}.

The expression for $G_\phi$ \refeq{grad prod ca} is obtained in a
similar way. First from \refeq{Phib} to get
$\nabla_{\phi_b}\Phi(\phi)=\nabla_{\phi_b}e^{i\Bcal(\phi)}=i\Pcal_b$
with $\Pcal_b$ from \refeq{Pcal} and then application of \refeq{vomv}
to \refeq{obj} returns
\beq[eq:vPbv]
(G_\phi)_{b,n} = -v_n^\dag(I_A\otimes P_b)v_n,\
b=1,\ldots,\nbsqr,\ n=1,\ldots,\n
\eeq
with $V=[v_1 \cdots v_N]\in\U(N)$ from \refeq{vomv}.  The rest of the
expressions in \refeq{Pcal} follow from ``vec'' operation. In
addition, using the fact that
$\sum_{b=1}^\nbsqr\vec{B}_b\vec{B}_b^\dag=I_\nbsqr$ together with the
invertibility of $\Kcal$, which implies that
$\Kcal\Kcal^\dag\geq\kappa I_\nbsqr,\kappa>0$, we get,
\beq[eq:ggphi lb]
G_\phi^TG_\phi
\geq
\kappa 
\sum_{b=1}^\nbsqr z_bz_b^\dag,\
z_b=
\left[
  \bea{c}
  v_1^\dag\left(\Is\otimes B_b\right)v_1
  \\
  \vdots
  \\
  v_N^\dag\left(\Is\otimes B_b\right)v_N
  \eea
  \right] \in\Rbf^{N}
\eeq
Though seemingly tractable, establishing the rank of $G_\phi$ from
this formulation is currently an open problem.

\section{Kinematic Critical Points}
\label{sec:kinJ}

The kinematic objective, gradient, and Hessian are,
\beq[eq:kinJ]
\bea{rcl}
J(\om) &=& \sum_{n=1}^N\cos\om_n \in[-N,N]
\\
\nabla_\om J(\om) &=& -\sin\om
\in\Rbf^N
\\
\nabla^2_\om J(\om) &=& -\diag(\cos\om)
\in\Rbf^{N\times N}
\eea
\eeq
The kinematic critical points are where
$\sin\om_n=0,\cos\om_n\in\{+1,-1\},n=1\:N$.  Let $p$ denote the number
of times $\cos\om_n=1$. Reordering the frequencies gives,
\beq[eq:cosp]
\cos\om_n=\left\{\bea{ll}
+1 & n=1\:p
\\
-1 & n=p+1\:N
\eea
\right.
\eeq
At $p=\{0,N\}$, $J(\om)=\{-N,+N\}$, the extrema objective values, and
for $p=1\:N-1$, $J(\om)=2p-N\in(-N,+N)$, the objective values in the
interior of the landscape. The kinematic critical points, denoted by
$\om\in\Rbf^N$, and the associated \emph{kinematic critical
  values}, $J(\om)$, can be obtained from standard trigonometry
producing the summary of their properties in \S\refsec{kincrit pts}.

\section{Extended landscape gradient}
\label{sec:dJphi}

To establish \refeq{atphiopt}, note first that from \refeq{gga} for a
given control $c$, at the optimal extended landscape variable $\Phi_\opt(c)$,
equivalently $\Phi(\phi_\opt(c))$, and from \refeq{vomv} and
\refeq{Phiopt},
\beq[eq:]
\bea{c}
\trace\left(W\otimes\Phi(\phi_\opt(c))\right)U(c)
= \sum_{n=1}^N\cos\om_n(c,\phi_\opt(c))
\\
+i\sin\om_n(c,\phi_\opt(c))
= \trace Q
\eea
\eeq
where $Q$ is the diagonal matrix of singular values of $\Gam$. Since
the singular values are real, it follows that the first result in
\refeq{atphiopt} holds, namely,
\beq[eq:sinomopt]
\sum_{n=1}^N\sin\om_n(c,\phi_\opt(c))=0
\eeq
The second result in \refeq{atphiopt} is established by using the
expressions in Appendix \refsec{dyngrad}, \ie, the gradient of the
objective with respect to an element of the extended landscape parameter
vector $\phi$ is,
\beq[eq:dJb]
\bea{rcl}
\nabla_bJ &=& \real\trace\nabla_b\Phi^\dag\Gam
=
\real\trace(i\Phi P_b)^\dag\Gam
\\
&=&
\imag\trace P_b\Phi^\dag\Gam
=
\imag\trace(T_{\rm right}^\dag P_b T_{\rm right})Q
\eea
\eeq
where the last expression arises by evaluating the gradient at the
optimal extended landscape variable \refeq{Phiopt}. Since $T_{\rm right}^\dag
P_b T_{\rm right}$ is Hermitian and $Q$ is the diagonal matrix of
singular values of $\Gam$, then the the trace is a real number. It
follows that $\nabla_bJ=0$ at the optimal extended landscape, and hence we get
the second result in \refeq{atphiopt}.

\section{Rank condition for $\SU(N)$}
\label{sec:sun}

From \refeq{rankGcphiopt}, if $\Uobj\in\SU(N)$ then $\det \Uobj=
1$. From the decomposition \refeq{vomv}, $\det
\Uobj=\exp\{i\sum_n\om_n\}=1$ iff $\sum_n\om_n=2\pi k$ for any integer
$k$, so a degree of freedom is lost from $N$ to $N-1$. To account for
this in the dynamic gradient set $\om = (\omb,\ \om_N)\in\Rbf^N$ with
$\omb=(\om_1,\ldots,\om_{N-1})\in\Rbf^{N-1}$ and $\om_N = 2\pi
k-\sum_{n=1}^{N-1}\om_n = 2\pi k-\un_{N-1}^T\omb$. The dynamic
gradient is then,
\beq[eq:grad sun]
\bea{rcl}
G_{c,\phi} &=& \bar{G}_{c,\phi}\left[ I_{N-1},\ -\!\un_{N-1} \right]
\in\Rbf^{LM+\nbsqr\times N}
\\
\bar{G}_{c,\phi} &=& \nabla_{c,\phi}\omb \in\Rbf^{LM+\nbsqr\times N-1}
\eea
\eeq
It follows that $\rank\bar{G}_{c,\phi}=N-1$ implies that $\rank
G_{c,\phi}=N-1$ which establishes \refeq{rankGcphiopt} for
$\Uobj\in\SU(N)$.

\section{Closed-system}
\label{sec:closed sys}

For a closed-system the extended landscape variable reduces to a scalar,
$\Phi=e^{i\phi}$, equivalently, $\nb=1$ and $N=\na$.  The extended
landscape objective and gradient are then,
\beq[eq:Jclosed]
\bea{rcl}
J(c,\phi) &=& \real~\trace~\Uobj(c,\phi)
\\
\nabla_\phi J(c,\phi) &=& \real(i)~\trace~\Uobj(c,\phi)
\\
\Uobj(c,\phi) &=& e^{i\phi}W^\dag U(c) \in\U(N)
\eea
\eeq
Using the spectral decomposition set $\Uobj=Ve^{i\Om}V^\dag$,
$V\in\U(N)$, $\Om=\diag(\om)$, then,
\beq[eq:dJphi]
\nabla_\phi J(c,\phi) = \real(i)\sum_{n=1}^Ne^{i\om_n}
= -\sum_n\sin\om_n
= \un_N^Tg(\om)
\eeq
where $g(\om)=-\sin\om\in\Rbf^\n$. At the optimal extended landscape
$\phiopt(c)=(\trace W^\dag U(c))^*/|\trace W^\dag U(c)|$, the
gradient matrix then becomes (dropping the $c,\phiopt(c)$ dependence
for ease of reading),
\beq[eq:grad closed]
G_{c,\phi} = \left[\bea{c}G_c \\ \un_N^T\eea\right]
\in\Rbf^{LM+1\times N}
\eeq
If $\rank G_c=r$, then $G_c$ has the singular value decomposition:
$G_c=U_c\left[\bea{cc}S_r&0\\0&0\eea\right]V_v^T$ with $U_c\in\U(LM)$,
$V_c\in\U(N)$, $S_r=\diag(s_1,\ldots,s_r)$ and $s_1=\norm{G_c}\geq s_2
\geq \cdots \geq s_r>0$. Since $\rank G_{c,\phi}=\rank
G_{c,\phi}^TG_{c,\phi}$, then,
\beq[eq:rankGG]
\bea{rcl}
G_{c,\phi}^TG_{c,\phi}
&=&
\left[\bea{cc}S_r^2+v_1v_1^T & v_1v_2^T\\v_2v_1^T & v_2v_2^T\eea\right]
\\
v &=& V_c^T\un_N = \left[\bea{c}v_1\\v_2\eea\right]\in\Rbf^\n
\eea
\eeq
with $v_1\in\Rbf^r$, $v_2\in\Rbf^{\n-r}$ and
$v^Tv=v_1^Tv_1+v_2^Tv_2=\n$.  Because $S_r$ is invertible, we can
apply the \emph{rank additivity lemma} associated with the \emph{Schur
  Complement} for an Hermitian matrix, \ie, for
$X=\left[\bea{cc}A&B\\B^\dag&C\eea\right]=X^\dag$, if $A$ is
nonsingular then,
\beq[eq:rank add]
\rank X = \rank A + \rank(C-B^\dag A^{-1}B)
\eeq
Apllying this to \refeq{rankGG} with $A=S_r^2+v_1v_1^T$, $B=v_1v_2^T$,
and $C=v_2v_2^T$ gives,
\beq[eq:rank add app]
\bea{rcl}
\rank G_{c,\phi}^TG_{c,\phi}
&=& \rank(S_r^2+v_1v_1^T)
\\
&+&
\rank v_2(1-v_1^T(S_r^2+v_1v_1^T)^{-1}v_1)v_2^T
\\
&=& r + \rank v_2v_2^T
\\
&=& r+1
\eea
\eeq
provided that $v_1^T(S_r^2+v_1v_1^T)^{-1}v_1 < 1$ which is always the case.
This establishes \refeq{rankGcphi closed}.

\end{appendix}

\end{document}